\documentclass[a4paper,fleqn,usenatbib]{mnras}

\usepackage{newtxtext,newtxmath}
 
\usepackage[T1]{fontenc}
\usepackage{ae,aecompl}

\usepackage{graphicx}	
\usepackage{amsmath}	
\usepackage{amssymb}	

\title[QPOs in compact binaries]{QPOs in compact binaries from small scale eruptions in an inner magnetized disk}

\author[N. Scepi et al.]{
Nicolas Scepi$^{1}$\thanks{E-mail: nisc6580@colorado.edu}
Mitchell C. Begelman,$^{1,2}$
and Jason Dexter$^{1,2}$
\\
$^{1}$JILA, University of Colorado and National Institute of Standards and Technology, 440 UCB, Boulder, CO 80309-0440, USA\\
$^{2}$Department of Astrophysical and Planetary Sciences, University of Colorado, 391 UCB, Boulder, CO 80309-0391, USA 
}

\date{Accepted XXX. Received YYY; in original form ZZZ}

\pubyear{2020}

\begin{document}
\label{firstpage}
\pagerange{\pageref{firstpage}--\pageref{lastpage}}
\maketitle

\begin{abstract}
Dwarf nov\ae\ (DNe) and low mass X-ray binaries (LMXBs) are compact binaries showing variability on time scales from years to less than seconds. Here, we focus on explaining part of the rapid fluctuations in DNe, following the framework of recent studies on the monthly eruptions of DNe that use a hybrid disk composed of an outer standard disk and an inner magnetized disk. We show that the ionization instability, that is responsible for the monthly eruptions of DNe, is also able to operate in the inner magnetized disk. Given the low density and the fast accretion time scale of the inner magnetized disk, the ionization instability generates small, rapid heating and cooling fronts propagating back and forth in the inner disk. This leads to quasi-periodic oscillations (QPOs) with a period of the order of $1000$ s. A strong prediction of our model is that these QPOs can only develop in quiescence or at the beginning/end of an outburst. We propose that these rapid fluctuations might explain a subclass of already observed QPOs in DNe as well as a, still to observe, subclass of QPOs in LMXBs. We also extrapolate to the possibility that the radiation pressure instability might be related to Type B QPOs in LMXBs. 
\end{abstract}




\section{Introduction}

Dwarf nov\ae\ (DNe) are accreting compact binary systems, in which a white dwarf rips matter from the surface of its solar type companion. The matter ultimately forms an accretion disk around the white dwarf. DNe are mostly known for their recurrent outbursts on month time scales, where their optical flux increases by factors of ~100 for approximately one week. The disk instability model (DIM, \citealt{lasota2001}) is currently the favored model to reproduce these outbursts. In the DIM, these outbursts are driven by a thermal-viscous instability in the accretion disk. The trigger for the instability is the sudden change of opacity in the hydrogen ionization regime, which forces the disk to transit between a hot, eruptive, bright state and a cold, quiescent, faint state \citep{hoshi1979}. In its traditional form, the DIM assumes that angular momentum transport is due to turbulence and uses an $\alpha$ prescription, where $\alpha$ is the ratio of the stress to the thermal pressure \citep{Shakura}. The traditional DIM requires the use of an ad hoc prescription with a higher $\alpha$ in eruption than in quiescence to reproduce the outburst of DNe (\citealt{cannizzo1988}, \citealt{smak1999}, \citealt{kotko2012}, \citealt{cannizzo2012}). However, this ad hoc prescription does not find theoretical support from MHD simulations (\citealt{hirose2014}, \citealt{coleman2016}, \citealt{scepi2018}). Recent work by \cite{scepi2018b} shows that the inclusion of a magnetic torque due to a magnetic outflow \citep{blandford1982} could help alleviate this issue, without requiring an ad hoc prescription for $\alpha$. Whether the magnetic field configuration is arbitrarily fixed \citep{scepi2019} or self-consistently evolved with the accretion flow \citep{scepi2020}, an inner strongly magnetized zone, where angular momentum is mostly removed by the outflow-driven torque, can help to produce a recurrence time and amplitude consistent with observations. \\ 

DNe also exhibit rapid fluctuations on time scales of a few tenths to a few thousand seconds. The rapid fluctuations are generally sorted between dwarf nov\ae\ oscillations (DNOs), long period dwarf nov\ae\ oscillations (lp-DNOs) and quasi-periodic oscillations (QPOs). While the presence of QPOs in DNe has been known for more than twenty years they have not received the same attention from the community as those in low mass X-ray binaries (LMXBs), though they share many common properties \citep{warner2004}. 

DNOs have a typical period of a few tens of seconds and have been detected in $\approx 50$ cataclysmic variables (CVs) \citep{warner2004}. The DNO period is positively correlated with the optical and extreme UV luminosity and so probably with the underlying mass accretion rate (\citealt{patterson1981}, \citealt{mauche1996}, \citealt{mauche2001}). DNOs are believed to be due to magnetically channeled accretion onto an equatorial belt located on the surface of the white dwarf (\citealt{paczynski1978},  \citealt{warner2002b}). This model requires a relatively high stellar magnetic field ($B\gtrsim10^4$ G) but not too high to avoid being in the regime of polars and intermediate polars (IPs), which are CVs where the accretion disk is entirely or partly disrupted by the magnetic pressure of the white dwarf. 

lp-DNOs are oscillations with a period around 4 times larger than DNOs that can coexist with the latter. They have been detected in around $\approx20$ CVs so far, though some DNOs may have been misclassified \citep{warner2004}. Unlike DNOs they do not show a correlation with luminosity \citep{warner2003}. It has been proposed that they are connected to the rotation of the white dwarf and might be due to accreted matter that has spread out on the surface of the white dwarf \citep{warner2004}.

QPOs are the oscillations with the longest period, ranging from a few hundred to a few thousand seconds. They are much less coherent than DNOs and lp-DNOs, with a typical $Q=(dP/dt)^{-1}\approx5-20$, and so are more difficult to detect by Fourier transform \citep{warner2004}. QPOs have been detected in $\approx50$ CVs so far but they may be more common than originally thought \citep{warner2004}. There are at least three types of QPOs: the DNO-related QPOs, the IP-related QPOs and the others. There may be more types of QPOs but we will keep to this already complicated nomenclature. 

DNO-related QPOs coexist with DNOs and have a period of $\approx15$ times the DNO period (typically a few hundred seconds). DNO-related QPOs also show a correlation with the luminosity, with the ratio of the DNO-related QPO period to the DNO period staying constant when the luminosity varies \citep{woudt2002}. Another relation between QPOs and DNOs is given by the existence of double DNOs, which are pairs of DNOs where the beat period corresponds to the period of a QPO observed at the same time (\citealt{hesser1974}, \citealt{marsh1998}, \citealt{steeghs2001}, \citealt{woudt2002}, \citealt{warner2003}). 

IP-related QPOs are QPOs with a period of $\approx1000$ s. They would be linked to the rotation of the primary by magnetospheric accretion onto a highly magnetized white dwarf \citep{patterson2002}. The high stability of the signal often seen in IPs might be reduced by a high accretion rate. Note that DNOs and lp-DNOs are not expected to coexist with IP-related QPOs since the intensity of the white dwarf's magnetic field is the varying parameter between these two. This exclusion is verified in the sample of \cite{warner2004}.

Finally, once we remove DNO-related QPOs and QPOs which are thought to be IP-related, there remains the other QPOs for which no model exists. These are the subjects of this article. We first present light curves obtained from the model of \cite{scepi2020}, in which rapid fluctuations have been identified (\S\ref{sec:framework}). We propose that these rapid fluctuations might correspond to this subclass of QPOs. We determine the physical origin of these rapid fluctuations and present an analytical model (\S\ref{sec:analytical_model}) giving the period of these rapid fluctuations as a function of the system parameters. We then compare our analytical model with the observations of DNe QPOs (\S\ref{sec:observations}). We also explore the possibility that our model applies to QPOs in LMXBs in   (\S\ref{sec:discussion_LMXBs}) before concluding.

\section{Framework}\label{sec:framework}
Since we use results coming from \cite{scepi2020}, we recall here the main features of their model. The novelty of the DIM used in \cite{scepi2020} is to self-consistently include magnetically outflow-driven accretion and transport of magnetic field. The turbulent and outflow-driven torques are included using prescriptions from 3D radiative shearing-box simulations from \cite{scepi2018b} and self-similar solutions from \cite{jacquemin2019}, respectively.  Heating is assumed to be due to turbulent-driven transport and the radiative cooling is computed using fits to the effective temperature in the different regimes of ionization from \cite{latter2012}. 

Magnetic field transport is modeled using very simple power laws to explore different behaviors. \cite{scepi2020} find that when the disk is able to advect magnetic flux inwards, it divides in two zones: an inner zone that accumulates magnetic flux, until it reaches an equilibrium magnetization, and an outer zone that empties of magnetic flux, if not replenished from the outer boundary. The hybrid disk structure is illustrated on Figure \ref{fig:schema_hybrid}.

In the inner zone, angular momentum is removed very efficiently by the magnetized outflow's torque giving a fast accretion time scale.  The effect of the inner magnetized zone is to truncate the outer $\alpha$-disk. Truncation of the inner disk in DNe and LMXBs is needed both observationally and theoretically (\citealt{livio1992}, \citealt{meyer1994}, \citealt{lasota1996}, \citealt{king1997}, \citealt{dubus2001}, \citealt{schreiber2003}, \citealt{yuan2014}, \citealt{zdziarski2020}) and an inner magnetized zone with a fast accretion time scale is an appealing way of providing one. The eruptions of DNe are due to the thermal-viscous instability in the outer disk as in the classical DIM. On the time scale of the eruptions of DNe, the inner zone is mostly passive. However, as we will see here, the inner zone exhibits rapid fluctuations that might be related to the observed rapid fluctuations of DNe. 

\subsection{Light curve and rapid fluctuations}
\begin{figure}
\includegraphics[width=\columnwidth]{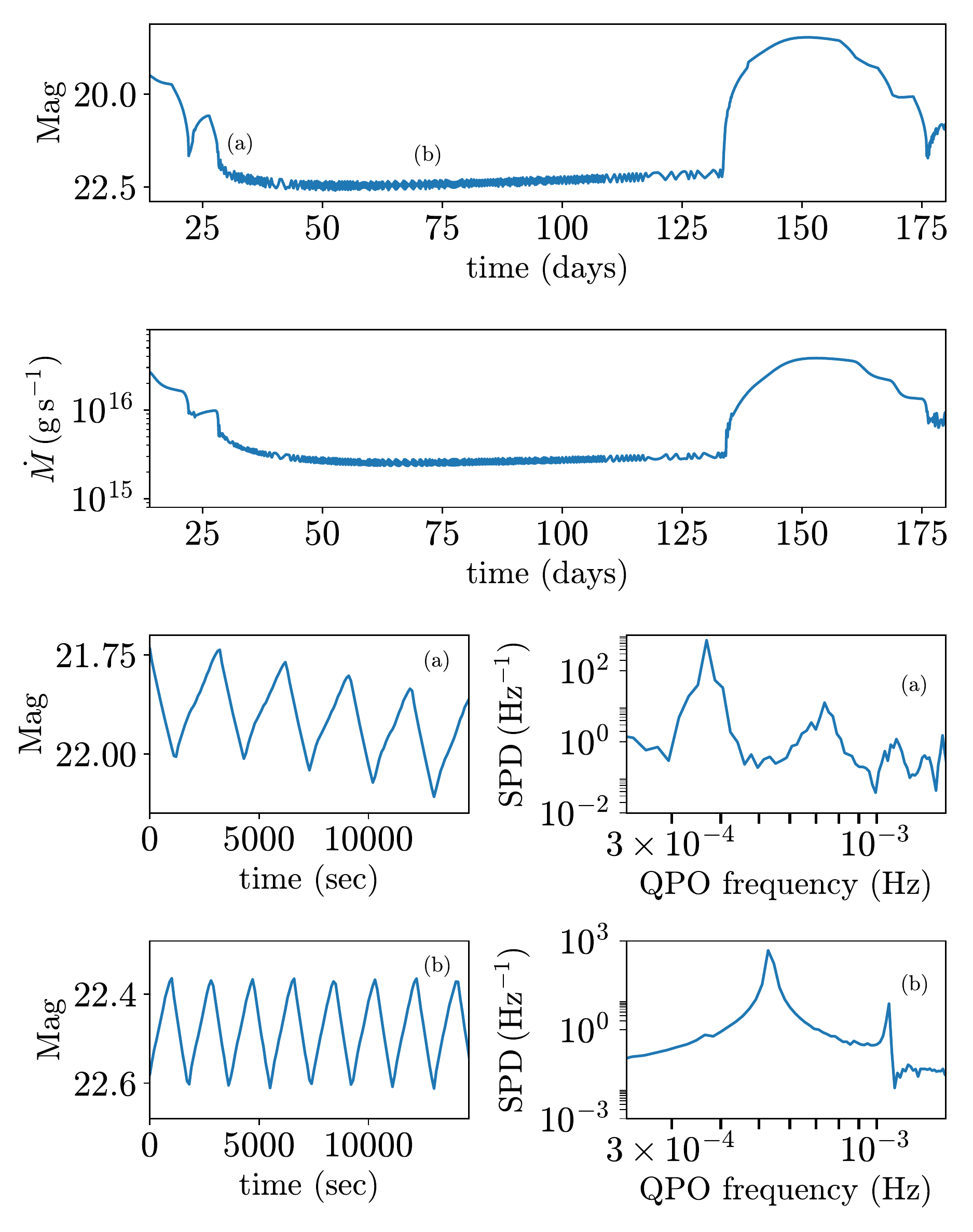}
\caption{First panel: light curve of DNe taken from \protect\cite{scepi2020} with the end of outburst marked as (a) and the deep quiescence marked as (b). Second panel : Mass accretion rate as a function of time. The initial uniform vertical magnetic field for this simulation is $10$ G and the external mass transfer rate is $\dot{M}_\mathrm{ext}=10^{16}\:\mathrm{g\:s^{-1}}$. Left bottom panels: Zoom-in on the light curves at times (a) and (b). Right bottom panels: Spectral power distribution of the signal (a) and (b).}
\label{fig:LC}
\end{figure}

Figure \ref{fig:LC} shows a light curve from \cite{scepi2020} where we can see the secular eruptions as well as small-scale, rapid fluctuations during quiescence. The shape of the light curve as well as the secular dynamics of the disk is discussed in \cite{scepi2020}. Here, we focus on the small-scale, rapid fluctuations that appear at the end of the outburst, last during the entire quiescent state and disappear at the beginning of the next outburst. We propose here that these rapid fluctuations may correspond to quasi-periodic oscillations observed in dwarf nov\ae\ and so we will refer to them as QPOs. 

We zoom in on two instants named (a) and (b), corresponding to the end of an eruption and the time of minimum luminosity in quiescence, respectively. The left bottom panels of Figure \ref{fig:LC} show the instants (a) and (b) in the temporal domain. We see rapid oscillations of $\gtrsim0.2$ magnitudes with a larger period in (a) than in (b). The bottom right panels show the spectral power density of the signals (a) and (b). The right (left) upper panel shows a broad (narrow) peak at the first harmonic as well as a peak at the second harmonic, which is fainter and as broad (slightly narrower) than the first harmonic. The frequencies of the fundamental peaks are, respectively, $3.4\times10^{-4}$ Hz and $5.4\times10^{-4}$Hz for (a) and (b). This sets the period of the QPOs in (a) and (b) to be 2941 s and 1852 s, respectively.\\

\begin{figure}
\includegraphics[width=90mm]{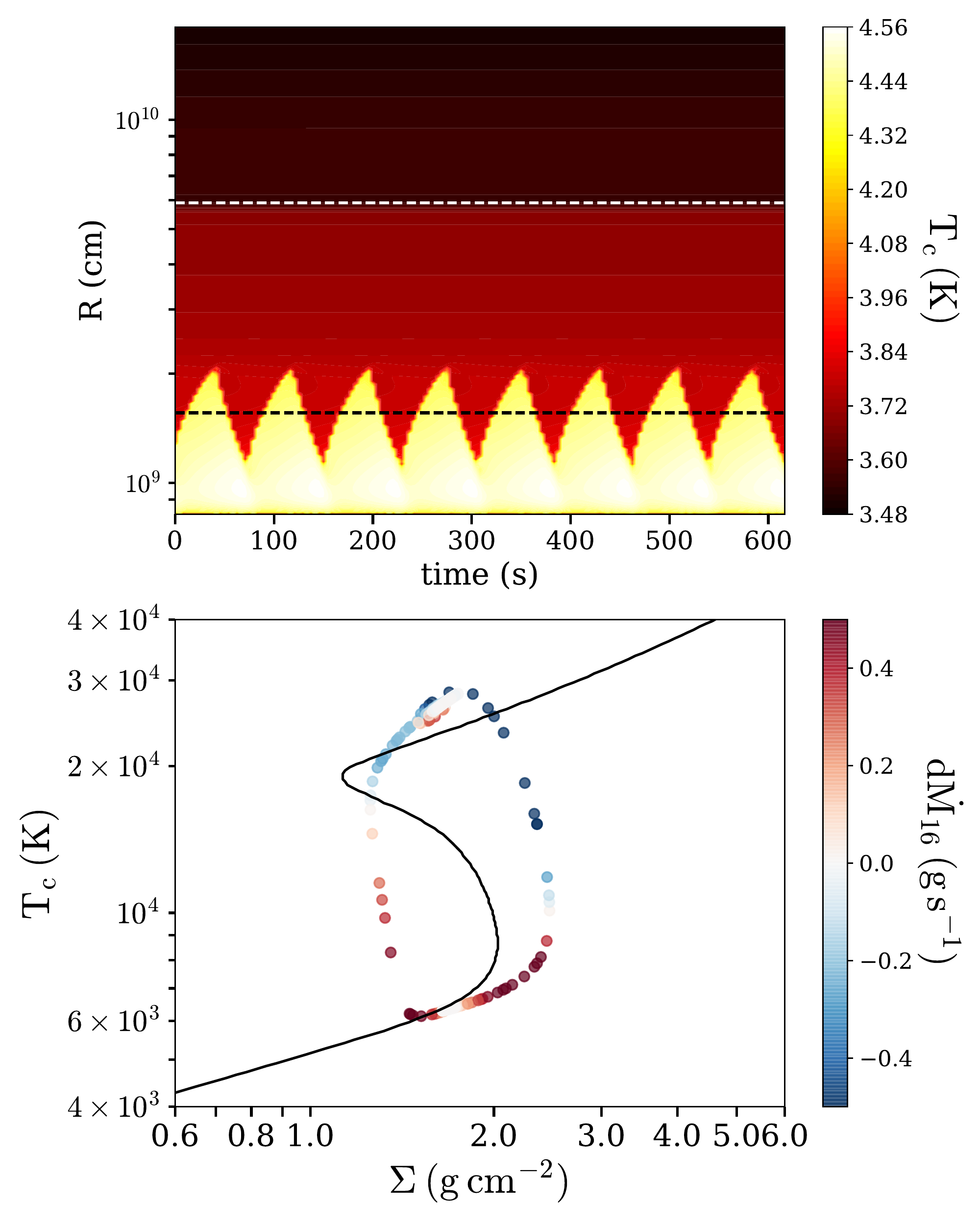}
\caption{Top panel: Temperature as a function of time and radius at instant (b) from Figure \ref{fig:LC}. The black dashed line represents our analytical estimate of $R_\mathrm{ioniz}$ from Eq. (\ref{eq:r_crit}). The white dashed line represents the analytical estimate of $R_\mathrm{tr}$ from Eq. (24) of \protect\cite{scepi2020}. Bottom panel: Temperature as a function of the surface density during instant (b) at a radius of $R=1.5\times10^9$ cm. The color gives the difference in mass accretion rates between two adjacent radii for $R=1.5\times10^9$ cm and $dR=2.4\times10^7$ cm, i.e. blue means that the density is decreasing and red that the density is increasing. The solid black curve represents an S-curve at $R=1.5\times10^9$ with $\alpha=1$ and no convection from \protect\cite{hameury1998}.}
\label{fig:scurve}
\end{figure}

To understand the origin of these fluctuations, we focus on time (b). The top panel of Figure \ref{fig:scurve} shows the temperature in the disk as a function of time and radius in quiescence at time (b). We see that the rapid fluctuations are due to the presence of small outbursts in the inner magnetized zone. 

The bottom panel of Figure \ref{fig:scurve} shows the evolution of the surface density, $\Sigma$, and the central temperature, $T_c$, in quiescence at a radius of $R=1.5\times10^9\:\mathrm{cm}$ during the same period. For reference, we draw the thermal equilibrium states as a function of $\Sigma$ and $T_c$ for a radius of $R=1.5\times10^9\:\mathrm{cm}$. The equilibrium states show a typical S-shape. The S-curve was computed with the DIM of \cite{hameury1998} with no convection and $\alpha=1$. The upper branch and lower branch are thermally and viscously stable while the middle branch is thermally and viscously unstable. The inflexion points of the S-curve are where the disk becomes thermally unstable due to hydrogen recombination (upper branch) or ionization (lower branch). Such an S-curve can also be computed in the outer disk. In the DIM, the eruptions are actually due to a viscous disk transiting between the two branches, with the surface density decreasing on the upper branch and increasing on the lower branch producing a hysteresis cycle. 

We see that the density and temperature evolution follows remarkably well the hot and cold stable branches of the S-curve in the inner disk. The points between the two stable branches are transitory points when a front is heating or cooling the disk. The color of the dots also gives $d\dot{M}\equiv\dot{M}(R+dR)-\dot{M}(R)$. For $d\dot{M}>0$ the surface density increases with time, while for $d\dot{M}<0$ it decreases with time. We see that on the hot branch, the density decreases while on the cold branch it increases. All of this suggests that the rapid fluctuations are an analog of the eruptions in the outer disk and are a consequence of the ionization instability operating in the inner outflow-driven disk. However, those eruptions are smaller than the eruptions in the outer viscous disk due to the difference of accretion time scales as we discuss in the following section. 

\subsection{Time scales}\label{sec:time_scale}
The recurrence time scale between eruptions corresponds to the time it takes to build enough mass in the quiescent state to reach the critical maximum surface density above which the disk is thermally unstable, $\Sigma_\mathrm{max}$. Alternatively, the decay time scale in eruption corresponds to the time it takes to empty the disk enough to reach the critical minimum surface density below which the disk is thermally unstable, $\Sigma_\mathrm{min}$. The propagation speed of the heating front is roughly given by the sound speed in the front while the propagation speed of the cooling front depends on the radius in the disk and disk parameters such as $\alpha$ (\citealt{lin1985}, \citealt{menou1999}). We will take, for simplicity, the time of propagation of the front to be the one of the heating front, which is 
\begin{equation}
t_\mathrm{front}=\frac{1}{\Omega}\left(\frac{R}{H}\right).
\end{equation}

In the outer, viscous disk, accretion happens on the viscous time scale, 
\begin{equation}
t_\mathrm{vis}=\frac{1}{\alpha\Omega}\left(\frac{R}{H}\right)^2,
\end{equation} 
where we use $\alpha=\mathrm{min}(15\times\beta^{-0.56}+0.03,1)$ according to \cite{scepi2018b}, with the plasma $\beta$ parameter defined as
\begin{equation}
\beta=\frac{8\pi P_\mathrm{mid}}{B_z^2},
\end{equation}
and $P_\mathrm{mid}$ the thermal pressure at the midplane.
In DNe, this gives a recurrence time scale of order a month and a decay time scale of order a week.\\

In the inner magnetized disk, the accretion is not viscous but due to the outflow-driven torque. The time scale of accretion is then given by 
\begin{equation}
t_\mathrm{acc}=\frac{1}{\Omega} \left(\frac{q}{\beta}\right)^{-1}\left(\frac{R}{H}\right),
\end{equation}
where $q=0.36\times\beta^{0.6}$ is the vertical torque due to a magnetized outflow \citep{jacquemin2019}. Hence, in a magnetized disk ($\beta\lesssim10^3$), the time scale of accretion is only a factor $(q/\beta)^{-1}$ larger than $t_\mathrm{front}$. For example, for $\beta=1$, $t_\mathrm{acc}\approx 3\times t_\mathrm{front}$. For comparison, in a viscous disk with $\beta=1$, $t_\mathrm{acc}=(R/H) \times t_\mathrm{front}$. 

\section{An analytical model for QPOs in DNe}\label{sec:analytical_model}

\subsection{Solutions for outflow-driven disk}
\begin{figure}
\includegraphics[width=90mm]{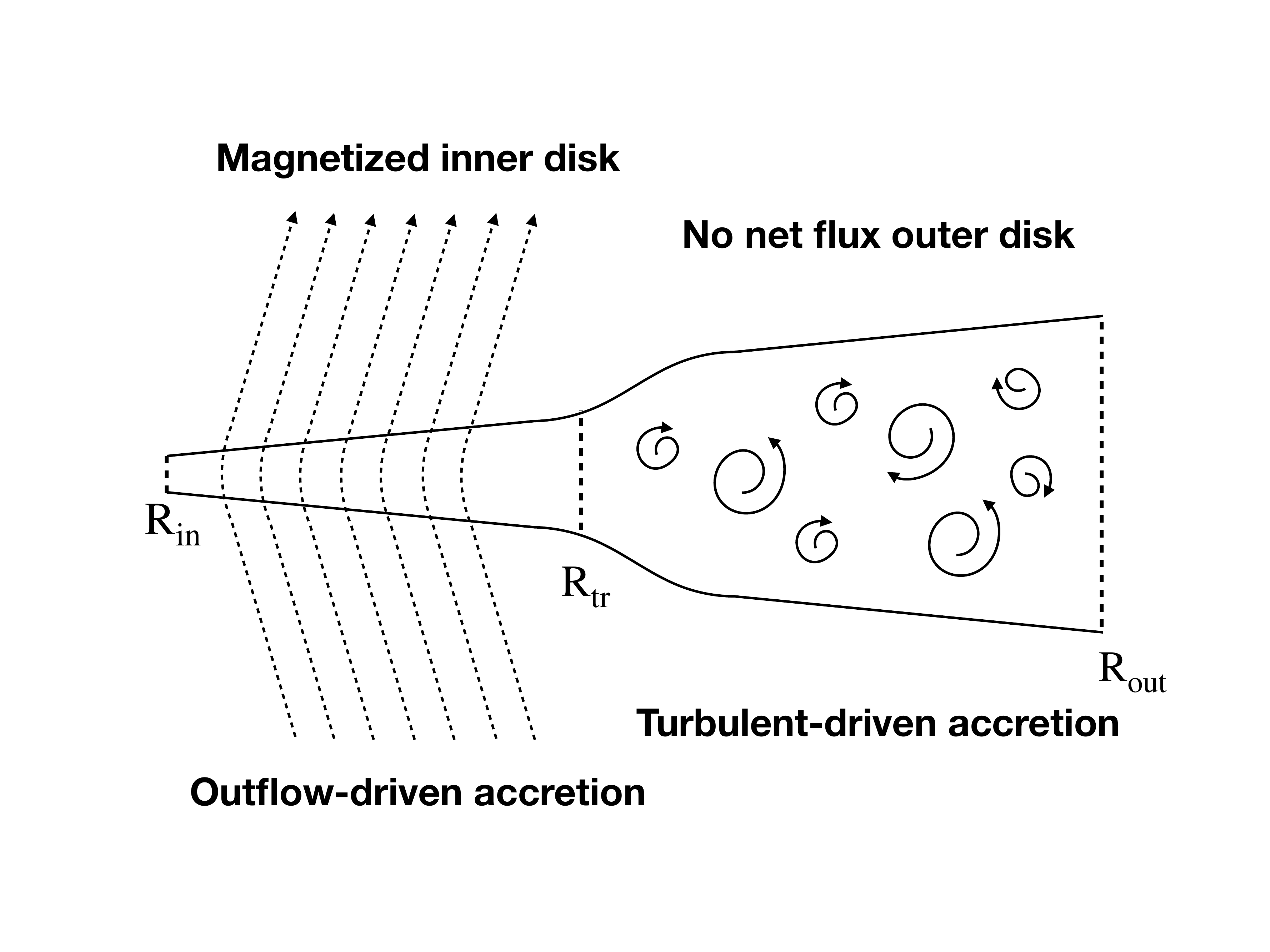}
\caption{Schematic of a hybrid disk.}
\label{fig:schema_hybrid}
\end{figure}

In order to provide an analytical model to predict the period of the QPOs, we will use the following framework for the disk. We assume a hybrid accretion disk composed of two zones as illustrated in Figure \ref{fig:schema_hybrid}.

The inner zone is highly magnetized. It has a constant $\beta=\beta_\mathrm{eq}$ as a function of radius and contains all the net vertical magnetic flux in the disk. Angular momentum is mostly removed by a large scale torque due to a magnetized outflow.   

The outer zone is weakly magnetized. Angular momentum is extracted by the magneto-rotational instability (MRI; \citealt{balbus1991}) turbulence and the disk is modeled as an $\alpha$-disk. Since it contains negligible net vertical magnetic flux, the turbulence is sustained by an MRI dynamo. 
 
The inner disk extends from the inner radius, $R_\mathrm{in}$, to the transition radius, $R_\mathrm{tr}$. The transition radius evolves as 
\begin{equation}\label{eq:r_tr}
R_\mathrm{tr}\propto \dot{M}^{-2/3}
\end{equation}
due to magnetic flux conservation in the hybrid disk \citep{scepi2020}. The outer disk extends from $R_\mathrm{tr}$ to the outer radius, $R_\mathrm{out}$.\\

Since the eruptions providing the QPOs appear in the inner magnetized disk, we will focus on this region of the disk only.

In the inner disk, the mass accretion rate can be expressed as 
\begin{equation}\label{eq:accretion}
\dot{M}\approx8\pi\frac{q}{\beta}P_\mathrm{mid}\frac{R}{\Omega},
\end{equation}
where $\Omega$ is the Keplerian frequency at $R$. Note that, for simplicity, we ignore the mass loss rate in the wind as it is expected to be small in DNe (\citealt{hoare1993}, \citealt{knigge1997}).

As in a classical $\alpha$-disk, the heating rate in the inner disk is due to turbulence and can be written as 
\begin{equation}\label{eq:heating}
Q^{+}=\alpha c_sH\Sigma \Omega^2,
\end{equation}
where $c_s$ is the sound speed, $H$ is the scale height of the disk, and $\Sigma$ is the surface density. However, note that in the inner disk the classical relation between the mass accretion rate and the heating rate does not hold \citep{scepi2018b}. 

The cooling rate can be expressed as 
\begin{equation}\label{eq:cooling}
Q^-=2\sigma_B T_\mathrm{eff}^4
\end{equation}
where the expression for the effective temperature $T_\mathrm{eff}$ depends on the optical depth in the disk. In the optically thick regime, we can use the flux diffusion approximation, and find
\begin{equation}\label{eq:diffusion_flux}
T_\mathrm{eff}^4=\frac{4T_c^4}{3\tau}
\end{equation}
where $T_c$ is the central temperature and $\tau$ is the optical depth in the disk, which can be approximately expressed as 
\begin{equation}\label{eq:tau}
\tau=\Sigma\kappa
\end{equation}
where $\kappa$ is the opacity. On the ionized, hot branch of dwarf nov\ae, we use
\begin{equation}\label{eq:kappa_hot}
\kappa_\mathrm{hot}=1.5\times10^{20}\rho_c T_c^{-2.5}
\end{equation}
following \cite{latter2012}, with $\rho_c$ the central density.

Assuming thermal equilibrium, i.e. $Q^+=Q^-$, we can now solve the system composed of (\ref{eq:accretion}), (\ref{eq:heating}), (\ref{eq:cooling}), (\ref{eq:diffusion_flux}), (\ref{eq:tau}), (\ref{eq:kappa_hot}) and find

\begin{gather}
T_c=4.0\times10^3\alpha^{2/15}\left(\frac{q}{\beta}\right)^{-2/5}M_1^{2/15}\dot{M}_{16}^{2/5}R_{10}^{-4/5}\:\mathrm{K}, \label{eq:OD_Tc} \\
\Sigma=0.2\alpha^{-1/15}\left(\frac{q}{\beta}\right)^{-4/5}M_1^{-1/15}\dot{M}_{16}^{4/5}R_{10}^{-3/5}\:\mathrm{g\:cm^{-2}},  \label{eq:OD_Sigma} \\
\frac{H}{R}=3.5\times10^{-3}\alpha^{1/15}\left(\frac{q}{\beta}\right)^{-1/5}M_1^{-7/15}\dot{M}_{16}^{1/5}R_{10}^{1/10}\label{eq:H_R},
\end{gather}
where $M_1$ is the mass of the primary in solar units, $\dot{M}_{16}=\dot{M}/10^{16}\:\mathrm{g\:s^{-1}}$, and $R_{10}=R/10^{10}\:\mathrm{cm}$.\\

\begin{figure}
\includegraphics[width=95mm,height=95mm]{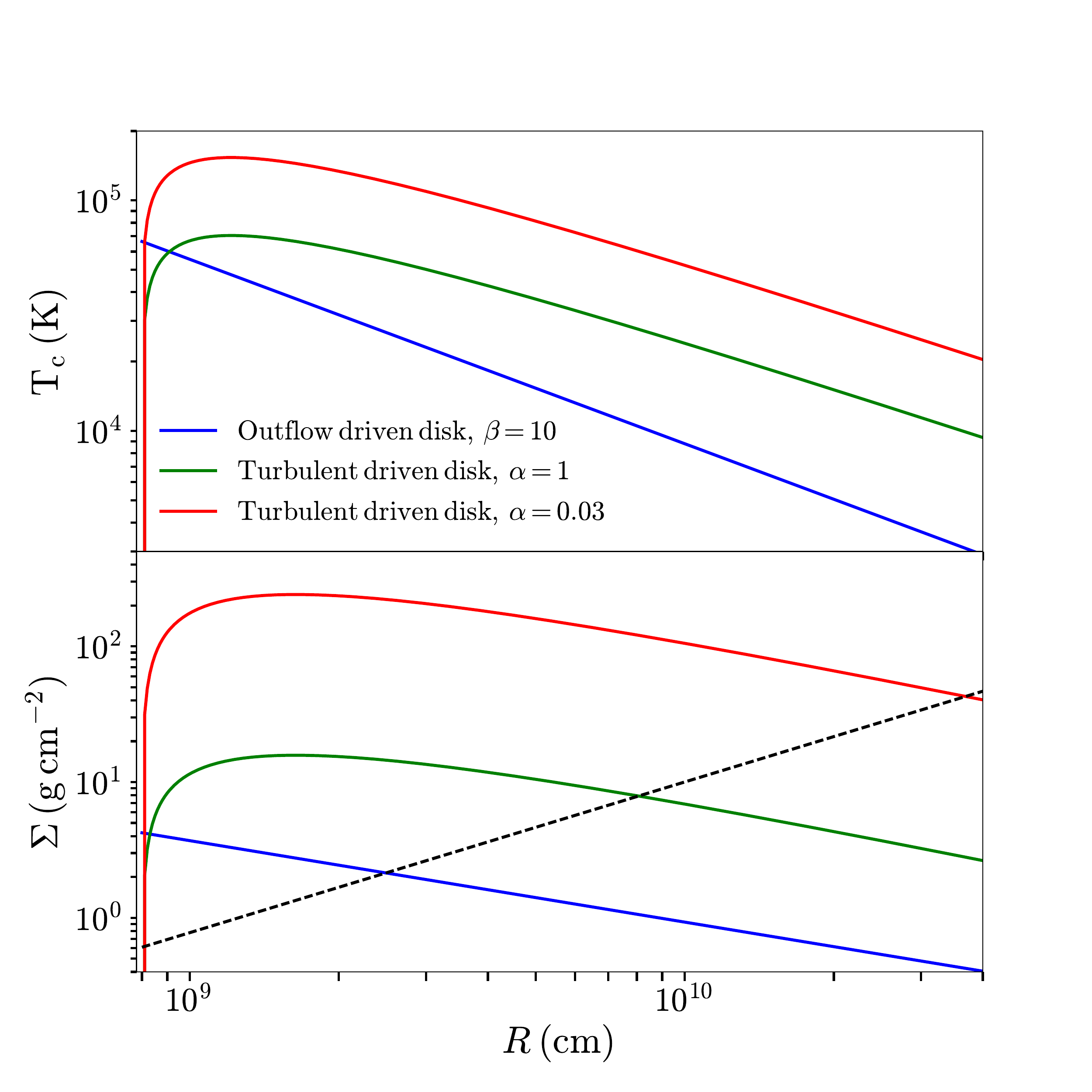}
\caption{Comparison of outflow-driven solutions with Shakura-Sunyaev solutions for $\dot{M}=10^{16}\:\mathrm{g\:cm^{-2}}$ and different values of $\beta$. The black dashed line indicates the critical surface density, $\Sigma_\mathrm{min}$, below which a hot disk cools back to the cold branch.}
\label{fig:OD_solution}
\end{figure}

We plot on Figure \ref{fig:OD_solution} the outflow-driven solutions in the hot state with, as a comparison, the classical $\alpha$-disk solutions using equation (\ref{eq:kappa_hot}) for the opacity and the prescription from \cite{scepi2018b} for $\alpha$ as a function of $\beta$. We note that changing the value of $\beta$ modifies $q/\beta$ and $\alpha$. For $\beta=10$, we have $\alpha=1$ while for $\beta\rightarrow \infty$ we have $\alpha=0.03$.

We see on Figure \ref{fig:OD_solution} that the outflow-driven solutions are less dense and colder than $\alpha$-disk solutions \citep{ferreira1995}. This is particularly true when we compare a weakly magnetized $\alpha$-disk (red curve) and a magnetized outflow-driven solution (blue curve). 

The critical surface density, $\Sigma_\mathrm{min}$, below which a hot disk becomes thermally unstable and cools back to the cold branch is also plotted as a black dashed line on Figure \ref{fig:OD_solution}. We take 
\begin{equation}\label{eq:sigma_min}
\Sigma_\mathrm{min}=8.3\alpha^{-0.77}M_1^{-0.37}R_{10}^{1.11}\:\mathrm{g\:cm^{-2}},
\end{equation}
in the optically thick regime \citep{hameury1998}. We see on Figure \ref{fig:OD_solution} that $\Sigma<\Sigma_\mathrm{min}$ at a radius of $\approx8\times10^{9}$ cm in the case of an $\alpha$-disk with $\alpha=1$ and at a radius of $\approx4\times10^{10}$ for $\alpha=0.03$. In the case of an outflow-driven disk, this happens at $\approx2.5\times10^9$ cm. In an outflow-driven disk, the ionization instability is triggered at smaller radii than in an $\alpha$-disk, for a given mass accretion rate, because of the lower density and temperature in the former.

\subsection{Estimate of $t_\mathrm{QPO}$}

Using (\ref{eq:OD_Sigma}) and (\ref{eq:sigma_min}), we then find that an outflow-driven disk becomes unstable at a radius given by
\begin{equation}\label{eq:r_crit}
R_\mathrm{ioniz}=1.1\times10^{9}\alpha^{0.41}\left(\frac{q}{\beta}\right)^{-0.47}M_1^{0.18}\dot{M}_{16}^{0.47}\:\mathrm{cm}. 
\end{equation}

The appearance of a cool spot in the disk triggers a cooling front. In the DIM, the heating front propagates on a time scale given by $\approx R/c_s$ \citep{menou1999} while the density evolves on the viscous time scale $t_\mathrm{vis}=\frac{1}{\Omega}(\frac{R}{H})^2$. Hence, there exists a clear separation between the thermal evolution and the viscous evolution. However, as we explained in \S\ref{sec:time_scale}, in the case of an outflow-driven disk, the accretion time scale is $\gtrsim R/c_s$ and $\ll t_\mathrm{vis}$. As a consequence, a heating front cannot propagate efficiently to the outer parts as the density drops below $\Sigma_\mathrm{min}$ on a time scale comparable to the front propagation time scale. This leads to fronts bouncing back and forth in the inner magnetized zone. We propose this to be the origin of the rapid fluctuations in the inner disk producing the QPOs.

Following this idea, we estimate the period of the QPO to be given roughly by 
\begin{equation}\label{eq:tqpo}
t_\mathrm{QPO}=\frac{2R_{\mathrm{ioniz}}}{c_s}.
\end{equation}
We note that this estimate is valid only when the variation of $R_{\mathrm{ioniz}}$ is of the order of $R_{\mathrm{ioniz}}$, which is always verified in our simulations (see for example Figure \ref{fig:scurve}).
Using (\ref{eq:H_R}), (\ref{eq:r_crit}) and (\ref{eq:tqpo}) we find 
\begin{equation}\label{eq:analytical_t}
t_\mathrm{QPO}=2.2\times10^{3}\alpha^{0.51}\left(\frac{q}{\beta}\right)^{-0.46}M_1^{0.19}\dot{M}_{16}^{0.46}\:\mathrm{s}.
\end{equation}

\subsubsection{Estimate of the amplitude}
The amplitude of the QPOs, in magnitudes, is given by  
\begin{equation}
\mathcal{A}=-2.5\log_{10}(1+\frac{\Delta L}{L_0})
\end{equation}
where $\Delta L$ is the bolometric luminosity emitted by the inner erupting disk and $L_0$ is the luminosity of the non-erupting part of the disk. 

In the hybrid disk, most (at least half in quiescence) of the luminosity is coming from the outer turbulent disk, due to the low radiative efficiency of the inner magnetized disk. Hence, we assume here that $L_0$ is given by 
\begin{align}
L_0=&\int_{R_\mathrm{tr}}^{R_\mathrm{out}}2\pi R\times2\sigma_B T_\mathrm{eff}^4dR\\
=&\int_{R_\mathrm{tr}}^{R_\mathrm{out}}\frac{3GM\dot{M}}{R^2}dR\\
\approx&\:4.0\times10^{31}\dot{M}_{16}M_1R_\mathrm{10,tr}^{-1}\:\mathrm{erg\:s^{-1}}
\end{align}
where $G$ is the gravitational constant. Note that we did not assume a zero-torque boundary condition at the inner boundary of the outer disk.

We estimate the luminosity emitted by the eruptive inner disk to be 
\begin{equation}
\Delta L= \int_{R_\mathrm{in}}^{R_\mathrm{ioniz}}2\pi R\times2\sigma_B T_\mathrm{eff}^4dR
\end{equation}
Using (\ref{eq:diffusion_flux}), (\ref{eq:tau}), (\ref{eq:kappa_hot}), (\ref{eq:OD_Tc}) and (\ref{eq:OD_Sigma}) we find 
\begin{equation}
\Delta L\approx 4.5\times10^{29}M_1^{17/30}\alpha^{16/15}\dot{M}^{6/5}(R_\mathrm{10,in}^{-9/10}-R_\mathrm{10,ioniz}^{-9/10}) \:\:\:\:\:\:\:\:\:\:\:\:\:\:\:\:\:\:\:\:\:\:\:\:\:\:\:\:\mathrm{erg\:s^{-1}}
\end{equation}
This gives 
\begin{equation}
\frac{\Delta L}{L_0}\approx1.1\times10^{-2}M_1^{-13/30}\alpha^{16/15}\left(\frac{q}{\beta}\right)^{-6/5}\dot{M}_\mathrm{16}^{1/5}\Delta R
\end{equation}
where $\Delta R=(R_\mathrm{10,in}^{-9/10}-R_\mathrm{10,ioniz}^{-9/10})R_\mathrm{10,tr}$.

To compare our analytical estimate with the QPO amplitude in Figure \ref{fig:scurve}, we adopt the following parameters: $R_\mathrm{10,in}=0.08$, $R_\mathrm{10,tr}=0.6$, $\beta=18$, $M_1=0.6$ and $\dot{M}_\mathrm{16}=0.25$. We find $|\mathcal{A}|\approx0.27$ mag in agreement with Figure \ref{fig:scurve} where we reported an amplitude of $\gtrsim0.2$ mag.

\subsubsection{Conditions for QPOs to occur}\label{sec:conditions}
Our model assumes that QPOs are due to the propagation of heating and cooling fronts in the inner magnetized zone. One strong requirement is that $R_\mathrm{in}<R_\mathrm{ioniz}<R_\mathrm{tr}<R_\mathrm{out}$. This ultimately puts a constraint on the range of mass accretion rates for which QPOs are expected to be present. 

The first condition is that $R_\mathrm{ioniz}>R_\mathrm{in}$. Using Eq. (\ref{eq:r_crit}) and a typical inner radius for a DNe of $8\times10^8\:\mathrm{cm}$, we find that 
\begin{equation}
\dot{M}_\mathrm{16,min}\approx0.5\left(\frac{q}{\beta}\right)M_1^{0.38}\alpha^{-0.87}
\end{equation} 
Since $\beta$ should typically range between 1 and $10^{3}$ for the inner disk to be considered as magnetized, the minimal mass accretion rate, $\dot{M}_\mathrm{16, min}$, for eruptions to occur in the disk is 
\begin{equation}
\dot{M}_\mathrm{16, min}\approx3\times10^{-2},
\end{equation}
where we used $M_1=0.8$. This means that systems with a very low accretion rate in quiescence should not be expected to show this type of QPO.\\

The second constraint is that $R_\mathrm{ioniz}<R_\mathrm{tr}$. \cite{scepi2020} showed that $R_\mathrm{tr}\propto\dot{M}^{-2/3}$ where the constant of proportionality depends on the amount of magnetic flux in the disk, $\Phi$. Moreover, for a DNe composed of a hybrid disk to have eruptions we must have $R_\mathrm{tr}<R_\mathrm{out}$ in eruption and in quiescence. Indeed, if there is no outer non-magnetized disk, there cannot be the day-to-week time scale eruptions, characteristic of DNe. This puts constraints on the amount of flux that is conceivable in a DNe. Using $\Phi=2\times10^{21}\:\mathrm{G\:cm^2}$ from the simulations in Figure \ref{fig:LC} as a reference case we find
\begin{equation}\label{eq:rtr_max}
R_\mathrm{10,tr}\approx2.7\times10^{-1}\dot{M}_\mathrm{16}^{-2/3}.
\end{equation}
Equating (\ref{eq:r_crit}) and (\ref{eq:rtr_max}), we find that the maximum accretion rate, $\dot{M}_\mathrm{16, max}$, for the eruptions to happen within the inner-magnetized disk is 
\begin{equation}
\dot{M}_\mathrm{16, max}\approx 2.2\left(\frac{q}{\beta}\right)^{0.41}M_1^{-0.16}\alpha^{-0.36}.
\end{equation}
Again since $\beta$ should typically range between 1 and $10^{3}$, this means, for $M_1=0.8$, that 
\begin{equation}
\dot{M}_\mathrm{16, max}\approx 1.5
\end{equation}
and so this type of QPOs should not be observed near the peak of outbursts.

\subsection{Comparison of the simulation and the analytical model}\label{sec:comparison_analytical_simu}
\begin{figure}
\includegraphics[width=90mm]{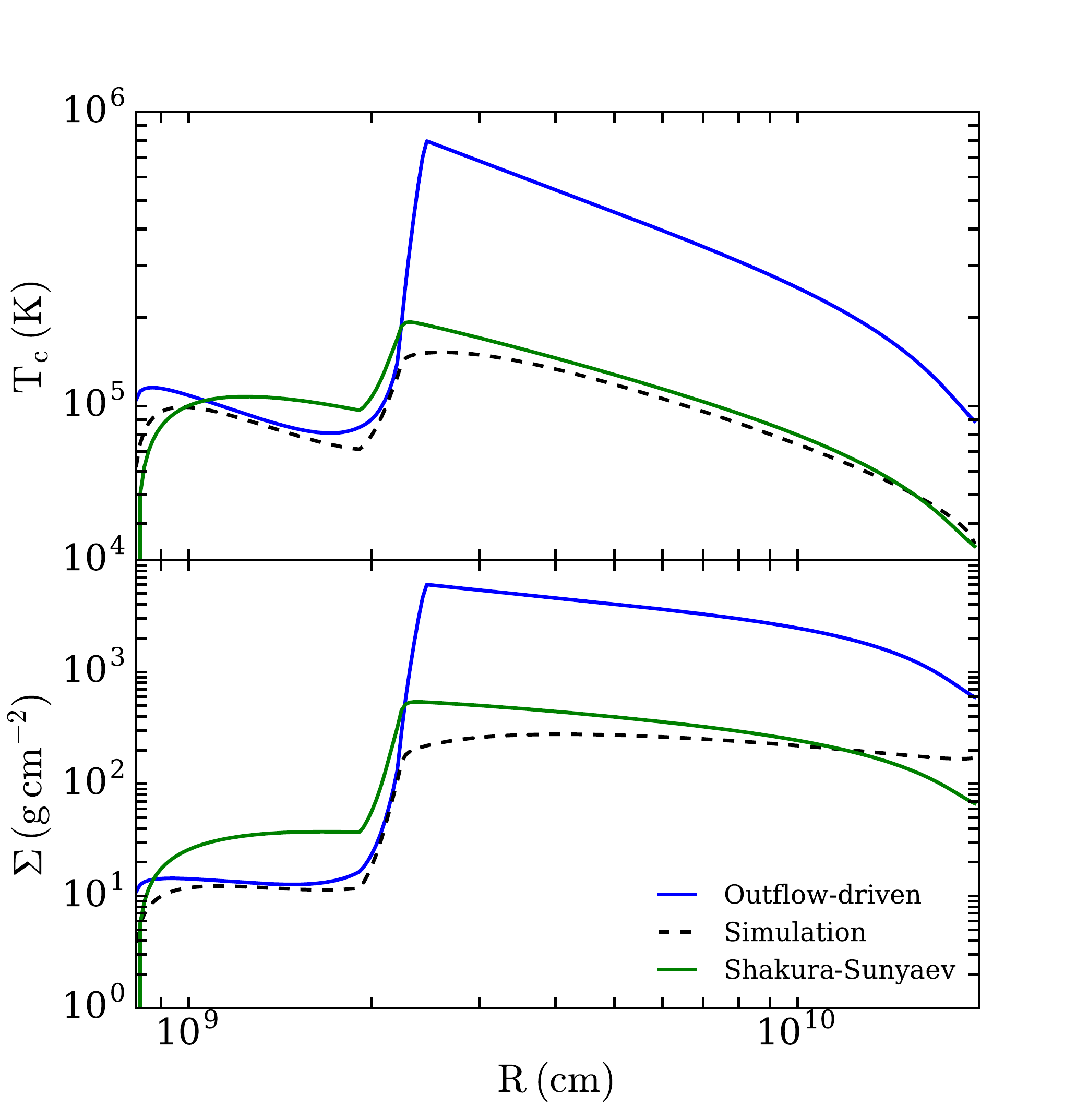}
\caption{Comparison of the outflow-driven solutions (blue solid line) and turbulent-driven solutions (green solid line) with a radial profile in eruption from \protect\cite{scepi2020} (black dashed line).}
\label{fig:OD_simu}
\end{figure}

We estimate the period of the QPOs by finding the radius at which the eruptive inner disk becomes quiescent. Hence, we need to verify that our analytical estimate of the disk structure in eruption is accurate. We plot on Figure \ref{fig:OD_simu} a comparison of our analytical solutions for the outflow-driven solutions and the turbulent-driven solutions with the results from the simulation of \cite{scepi2020}. The black-dashed line shows a radial profile of the disk in eruption from \cite{scepi2020}. To plot the analytical outflow-driven solution we used Eq. (\ref{eq:OD_Tc}) and (\ref{eq:OD_Sigma}) with the parameters taken from the simulation of \cite{scepi2020}. We see that, in eruption, the inner region is well approximated by a steady state outflow-driven disk while the outer region is well approximated by a steady state turbulent-driven disk. \\

\begin{figure}
\includegraphics[width=80mm]{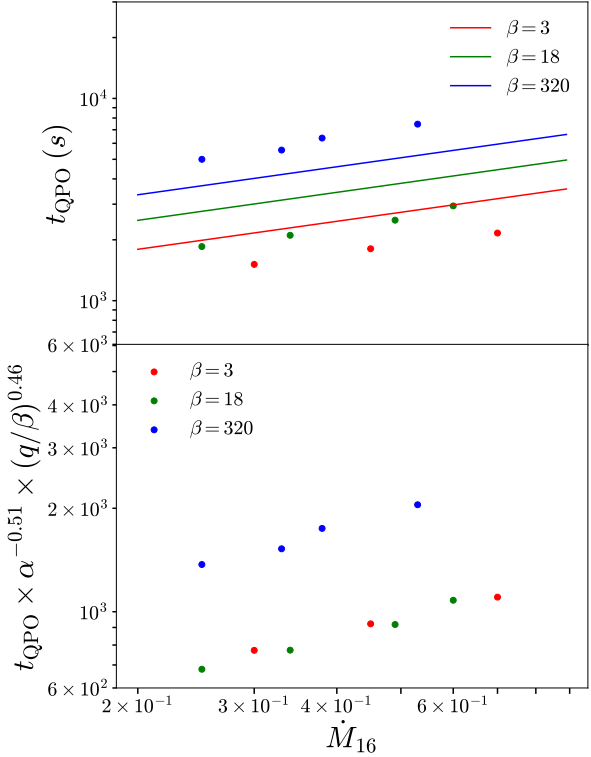}
\caption{Top panel: Comparison of the QPO periods from our analytical model (solid lines) with the measured QPO periods from the simulations of \protect\cite{scepi2020} for several values of $\beta$ as a function of $\dot{M}$ (colored dots). Bottom panel: Measured QPO periods from the simulations of \protect\cite{scepi2020} normalized by their expected analytical dependence on $\beta$ as a function of $\dot{M}$. }
\label{fig:scaling}
\end{figure}

Figure \ref{fig:scaling} shows the periods of the QPOs found in the simulations using the model of \cite{scepi2020} compared to the analytical predictions derived here. We performed three different simulations with $\beta=3$ (red dots), $\beta=18$ (green dots) and $\beta=320$ (blue dots) and measured the QPOs' period at several moments in each simulation. We see in the top panel of Figure \ref{fig:scaling} that the scaling with the mass accretion rate is the same for our analytical model and the simulations. Moreover, at low $\beta$, our model tends to overestimate the period by a factor $\approx1.4$ the period of the QPOs while in the high $\beta$ case we underestimate the period by the same factor.  

The bottom panel of Figure \ref{fig:scaling} shows the period of the QPOs from the simulations normalized by the expected dependence on $\beta$. We see that the $\beta=3$ and $\beta=18$ simulations scale well with our analytical expectation but not the $\beta=320$ simulation. This might be due to the fact that we approximate the angular momentum transfer as being due to the outflow only in our analytical model. This is a good approximation at low $\beta$ but seems to break down when we approach $\beta\approx1000$ where turbulent transport is no longer negligible. Taking into account both turbulent and outflow-driven transport is important as the time scale on which density evolves is a combination of a viscous time scale and the outflow-driven accretion rate time scale. 

\section{Discussion}
\subsection{Comparison to observations}\label{sec:observations}

Our aim is to provide a model for QPOs of DNe that are not DNO-related, lpDNO-related or suspected to be IP-related. We select objects from the Table 1 of \cite{warner2004} that respect these criteria. We removed from Table 1 of \cite{warner2004} all QPOs showing a period ratio of $\approx 15$ with DNOs (the DNO-related QPOs) and all QPOs from certified or suspected IPs. Out of the 76 objects in Table 1 of \cite{warner2004} we only have 9 objects potentially showing the type of QPOs we are interested in. We give in Table \ref{tab:obs} the name of the objects, their subclass, the period of the observed QPOs, their amplitude, their wavelength and the state in which they appeared. A striking feature of the remaining QPOs is that out of the 7 confirmed detections 6 of them are made during quiescence or at the end of an outburst. This was not a criterion of selection. Note that the two last objects, where QPOs are seen in outburst, are suggestions of detection by \cite{warner2003} and \cite{warner2002b} but not confirmed detections.\\

\begin{table*}
\begin{centering}
\begin{tabular}{ c c c c c c c }
\hline
 Name & $t_\mathrm{QPO}$ & State of appearance & Amplitude & Wavelength & Subtype & Reference \\
\hline
VZ Pyx & $\approx3000$ s & End of outburst & $\approx0.2$ mag & Optical, X-ray & SU Uma & [1][2]\\
WX Hyi & $\approx1140,1560$ s & 1 mag below outburst & $\approx10\%$ & Optical & SU Uma & [3]\\
OY Car & $\approx2193,3510$ s & Quiescence & $\approx20-40\%$ & X-ray & SU Uma & [4][5]\\
RAT J1953+1859 & $\approx1200$ s & Quiescence & $\approx$ 0.3 mag  & Optical & SU Uma & [6]\\
VW Hyi & $\approx1282$ s & Quiescence & $\approx$0.05 mag & Optical & SU Uma & [7]\\
SS Aur & $\approx1500$ s & Quiescence & $\approx0.2-0.4$ mag & Optical & U Gem & [8]\\
HX Peg & $\approx1800$ s & Always & $\approx$ 0.2 mag & Optical & Z Cam & [9]\\
V2051 Oph & $\approx1800$ s & Outburst (suggested) & x & Optical & SU Uma & [10]\\
RX And & $\approx$1000 s & Outburst (weak evidence) & x & Optical & Z Cam & [2] \\

\hline
\end{tabular}
\caption{Observational evidence of QPOs. References: [1] \protect\cite{remillard1994}, [2] \protect\cite{warner2003}, [3] \protect\cite{kuulkers1991}, [4] \protect\cite{ramsay2001}, [5] \protect\cite{hakala2004}, [6] \protect\cite{ramsay2009}, [7] \protect\cite{blackman2010}, [8] \protect\cite{tovmassian1988}, [9] \protect\cite{warner2003}, [10] \protect\cite{warner2002b}.
}
\label{tab:obs}
\end{centering}
\end{table*}

\begin{table*}
\begin{centering}
\begin{tabular}{ c | c | c | c | c | c c c }
\hline
Name & $M_1$ & $\dot{M}_\mathrm{16}$ & Reference $\dot{M}_\mathrm{16}$ & $t_\mathrm{QPO\:obs}\:(s)$ & \multicolumn{3}{c}{$t_\mathrm{QPO\:theo}\:(s)\:\mathrm{for\:different\:\beta}$} \\ & & & & & $\beta=1$ & $\beta=10$ & $\beta=100$ \\
\hline
VZ Pyx & $0.80\pm0.16$ & $\approx0.3$ & [1] & $\approx3000$ & 1939 & 2962 & 4524 \\
WX Hyi & $0.90\pm0.30$ & $0.5-1$ & [1] & $\approx1140,1560$ & 2508-3450 & $ 3831-5270$ & $ 5853-8051$ \\
OY Car & $0.685\pm0.011$ & 0.06-0.3 & [2][3] & $\approx2193,3510$ & 898-1883 & 1372-2876 & 2095-4393  \\
RAT J1953+1859 & x & x & x & $\approx1200$ & x & x & x \\
VW Hyi & $0.67\pm0.22$ & 0.08 & [4] & $\approx1282$ & 1021 & 1559 & 3674 \\
SS Aur & $1.08\pm0.40$ & 0.06 & [2] & $\approx1500$ & 979 & 1496 & 2285 \\
HX Peg & $0.75\pm0.15$ & 10-17 (Outburst) & [1] & $\approx1800$ & $\dot{M}_\mathrm{16}\:\mathrm{too\:high}$ & $\dot{M}_\mathrm{16}\:\mathrm{too\:high}$ & $\dot{M}_\mathrm{16}\:\mathrm{too\:high}$ \\
V2051 Oph & $0.78\pm0.06$ & 4 & [5] & $\approx1800$ & $\dot{M}_\mathrm{16}\:\mathrm{too\:high}$ & $\dot{M}_\mathrm{16}\:\mathrm{too\:high}$ & $\dot{M}_\mathrm{16}\:\mathrm{too\:high}$  \\
RX And & $1.14\pm0.33$ & 5-10 & [1] & $\approx$1000 & $\dot{M}_\mathrm{16}\:\mathrm{too\:high}$ & $\dot{M}_\mathrm{16}\:\mathrm{too\:high}$ & $\dot{M}_\mathrm{16}\:\mathrm{too\:high}$ \\
\hline
\end{tabular}
\caption{Comparison of theoretical and observed QPOs periods. Mass estimates come from \protect\cite{baptista1998}, \protect\cite{wood1989}, \protect\cite{dubus2018} and references therein. References for mass transfer estimates: [1] \protect\cite{dubus2018}, [2] \protect\cite{urban2006}, [3] \protect\cite{wood1989}, [4] \protect\cite{pandel2003}, [5] \protect\cite{baptista2007}. For VZ Pyx, we estimate $\dot{M}$ at the end of the outburst to be $\dot{M}$ at maximum from \protect\cite{dubus2018} divided by 5 as in our simulations. For WX Hyi, we use the estimate of $\dot{M}$ in outburst from \protect\cite{dubus2018}, that uses the new GAIA parallax, divided by 2.5 to take into account that the QPO is seen 1 magnitude below outburst.
}
\label{tab:comparison}
\end{centering}
\end{table*}

In Table \ref{tab:comparison}, we give the estimated mass of the primary and mass accretion rate from the literature, where these exist, as well as the analytical estimate of the QPO period from eq. (\ref{eq:analytical_t}) for three different values of $\beta$. We see that our model reproduces very well the expected order of magnitude of the QPO periods and in certain cases like VZ Pyx gives values that are very close to the observed values. Another strength of our model is the prediction that these QPOs should only appear in quiescence and at the end or beginning of an outburst. As we emphasized before, this was not a criterion of selection and yet 6 of the 7 confirmed detections were made in quiescence or at the end of an outburst. For the one remaining object showing a confirmed QPO in outburst and in quiescence our model cannot provide an explanation as the mass accretion rate in eruption is expected to be too high and the period of the QPO to change between outburst and quiescence.  \\

Finally, we focused here on the emission coming from the disk. Given the radii at which the disk becomes unstable, QPOs, in our model, would be emitted in the near UV in agreement with observations. However, some QPOs are also detected in X-rays (\citealt{remillard1994}, \citealt{ramsay2001}, \citealt{hakala2004}). X-rays in DNe come from accretion onto a boundary layer between the inner radius of the disk and the surface of the white dwarf (see \citealt{mukai2017} for a review). One natural expectation is that the small fluctuations of the mass accretion rate in the inner zone during the eruptions leading to QPOs (see second panel of Figure \ref{fig:LC}) provide a modulation in the amount of X-rays produced in the boundary layer. We did not try here to compute X-ray light curves and leave it to future work. We note that, in \cite{hakala2004}, the QPOs of OY Car, which were observed simultaneously, were proposed to originate from the white dwarf's rotation and an interference of the white dwarf's rotation with the orbital period of the system. We propose, in the same way, that the lower period QPO could originate from the ionization instability in the inner disk, and not the white dwarf, while the higher period QPO could still originate from the interference with the orbital period.\\

Our model depends on parameters such as $q$ and $\beta$, which are very poorly constrained observationally. Fortunately, the QPO period varies only as $\beta^{0.18}$ and so our estimates do not depend critically on a good knowledge of the magnetization unlike our estimate of the  QPO's amplitude, which varies as $\beta^{0.48}$. In any case, our model relies on a magnetized disk and so $\beta$ can only realistically range between $\approx1$ and $\lesssim10^3$. Hence, our uncertainty on the period due to a lack of information on the magnetization is only a factor of $\approx3.5$, so that we at least have confidence on the order of magnitude of the QPO period. The amplitude, however, can vary by an order of magnitude, over the range of magnetization considered here, going from a few percent for $\beta=1$ to a few tens of percent for $\beta=10^3$. This can make the difference between a detectable QPO and an undetectable one.

This is critical as our model has to explain not only the existence of QPOs in systems in which they are observed but also their absence in systems in which they are not observed. We estimated in \S\ref{sec:conditions} the conditions for QPOs to occur and found that the maximum range of accretion rate was given by $0.03<\dot{M}_\mathrm{16}<0.8$. However, this range is narrower for smaller $\beta$ and can be as narrow as $0.2<\dot{M}_\mathrm{16}<1.5$ for $\beta=1$. This narrow range of luminosity in which one can observe the QPOs coupled to the small amplitudes at $\beta=1$ converge to make the detection of these QPOs more challenging. To conclude, we suggest that the small amplitudes at large magnetizations of the QPOs as well as their restricted window of appearance (at the beginning and end of oubursts and during quiescence) could contribute to their low number of detections. We also should keep in mind that there exists an observational bias towards observing QPOs in outbursting DNe and the small overall number of detected QPOs in quiescence might be due partly to the fact that we have not been looking for them \citep{pretorius2006}.

\subsection{Limitations of the model}
In our model (analytical and numerical), we assume that specific angular momentum is carried away by a magnetized outflow but do not take into account the mass loss in the disk. We believe this to be a reasonable approximation as the mass loss in DNe is believed to be only a few percent of the accretion rate (\citealt{hoare1993}, \citealt{knigge1997}). Energy carried in the outflow, however, is unconstrained and might account for a significant part of the energy deposited through turbulent heating in the disk. 

We omit in our analytical model the treatment of magnetic field transport. In the simulation, the fluctuations in temperature and density during QPO eruptions lead to fluctuations in the local value of the magnetic field, since the magnetization is the fixed paramater. We performed a simulation in quiescence with a fixed magnetic field configuration and checked that the QPOs are still present in this case and that the period changed only by 30$\%$, justifying our neglect of magnetic field transport in the analytical model. 

We also used a simplified treatment of the thermodynamics. The prescription of \cite{latter2012} based on \cite{faulkner1983} is known to produce DNe-sized eruptions even in a disk with a constant $\alpha$. It is then possible that we overestimate the amplitude of the eruptions affecting the period of the QPOs. However, even more crude is our treatment of the front's physics. We do not treat here the advection of energy through a front or the adiabatic cooling/heating as the gas is expanding/compressing. We performed an additional simulation including these terms in the energy equation and found that the QPO period changes only by 24$\%$. We also use a simplified prescription for the heating compared to more complete DIM as in \cite{hameury1998}. Note that there is no correct analytical model to treat the thermodynamics of the front and \cite{hameury1998} estimate that the cooling fluxes near the front are not known with a better precision than 50$\%$ in their DIM. 
 
\subsection{Expectation of QPOs in 3D global simulations}
We wish to give here an idea of what features related to these QPOs could be observed in 3D global simulations. We estimate the time scale of these QPOs to be, according to (\ref{eq:tqpo}), equal to $\frac{2}{\Omega}\left(\frac{R}{H}\right)$. DNe are well-modeled by thin disks with $H/R\approx 10^{-2}-10^{-3}$. If the front propagates between the inner boundary and $\approx3$ times the inner boundary, these QPOs would then require a simulation to be evolved for at least $\approx1000$ inner orbital periods to see one front developing. This is within the current capacities of 3D radiative GRMHD simulations such as the one of \cite{jiang2020} who studied (unrelated) opacity-driven cycles on time scales of $\approx3000$ inner orbital periods while marginally resolving the MRI in a disk with $H/R\gtrsim10^{-2}$. 

\subsection{Application to LMXBs}\label{sec:discussion_LMXBs}
\subsubsection{QPOs in LMXBs}

QPOs in LMXBs are generally sorted between low-frequency QPOs with frequencies ranging from a few mHz to a few tenths of a Hz and high frequency QPOs with frequencies $\gtrsim 60$ Hz (\citealt{motta2016}, \citealt{ingram2020}). We will focus here on low frequency QPOs which are themselves sorted into three types: A, B and C. Type A QPOs are seen in the high soft state and have a frequency of 6-8 Hz \citep{motta2016}. Type B QPOs are seen at the transition between the hard intermediate state and soft high state of LMXBs and at the transition between the soft transition state and the hard intermediate state. They have a frequency around 6 Hz or 1-3 Hz \citep{motta2016}. They have been suggested to be related to a quasi-periodic heating of the disk or to changes at the base of the jet (\citealt{fender2009}, \citealt{miller2012}, \citealt{stevens2016}, \citealt{russell2019}).  Finally, Type C QPOs are seen in almost every spectral state and have frequencies ranging from a few mHz to about 10 Hz \citep{motta2016}. 

Since Type A and Type C QPOs are seen in the high soft state, it is unlikely that they could be explained by our model that requires an inner magnetized zone, inconsistent with the spectral shape seen in the high soft state. However, Type B QPOs are observed when the inner magnetized zone is assumed to disappear or reappear. Plus, the propagation of a front is likely to affect the disk and the jet, and so Type B QPOs are a good candidate for our model.

\subsubsection{Ionization instability}
To compare the QPO frequencies with our model, we rewrite (\ref{eq:analytical_t}) as
\begin{equation}
\mathrm{\nu_{QPO}}=7.1\times10^{-5}\alpha^{-0.51}\left(\frac{q}{\beta}\right)^{0.46}M_1^{-0.68}\dot{m}^{-0.46}\:\mathrm{Hz}
\end{equation}
where $\dot{m}=\dot{M}c^2/10L_\mathrm{Edd}$ and $L_\mathrm{Edd}$ is the Eddington luminosity . 

We can also rewrite (\ref{eq:r_crit}) as 
\begin{equation}\label{eq:rioniz}
\mathrm{r_{ioniz}}=5.1\times10^4\alpha^{0.41}\left(\frac{q}{\beta}\right)^{-0.47}M_1^{-0.35}\dot{m}^{0.47}
\end{equation}
where $r=R/r_g$ with $r_g=GM/c^2$.
One requirement of our model is that $\mathrm{r_{ioniz}}$ must be located in the inner magnetized region, or the ``truncated" region of LMXBs. We use the data of two different groups compiled and analyzed in \cite{tetarenko2020} that estimate observationally the transition radius in GX339-4. The study from \cite{marcel2019} fits the spectral evolution of GX339-4 with a hybrid jet emitting disk-standard disk model while \cite{garcia2015} and \cite{wang2018} use X-ray reflection spectroscopy data to retrieve the values of $\dot{M}_\mathrm{in}$ and $r_\mathrm{tr}$ during an outburst. The two studies agree that during a whole cycle $r_\mathrm{tr}<10^3$. More specifically, we use the results of \cite{marcel2019} to estimate $r_\mathrm{tr}$ throughout the cycle and find 
\begin{equation}\label{eq:rtr_marcel}
r_\mathrm{tr}\approx10^3q^{1/2}M_1^{-5/3}\dot{m}^{-2/3}.
\end{equation}
Note that this is an estimate for a particular outburst of GX339-4 and should change between objects as a function of the magnetic flux present in the disk. Nonetheless this gives an estimate of the maximum accretion rate, $\dot{m}_\mathrm{max}$, for QPOs to be produced,
\begin{equation}
\dot{m}_\mathrm{max}\approx3.2\times10^{-2}\alpha^{-0.36}q^{0.44}\left(\frac{q}{\beta}\right)^{0.41}M_1^{-1.16}.
\end{equation}

As in DNe, we would not expect those QPOs near the peak of an eruption of LMXBs. However, they could be present at the end of an outburst and in quiescence with typical frequencies of $\approx0.2$ mHz for $\dot{m}\approx10^{-3}$, $\beta=10$ and $M_1=5$. This is lower than the low end of observed low frequency QPOs in neutron star LMXBs. However, we note that the detections of sub-mHz QPOs is often prohibited by short observational windows. In quiescence the frequency would increase but then the system is fainter, making observations of such QPOs challenging. 

\subsubsection{Radiation pressure instability}
Another well-known instability that should develop in the inner regions of LMXBs disks is the radiation pressure instability (\citealt{Shakura}, \citealt{shakura1976}, \citealt{piran1978}). The radiation instability also produces an S-curve with the low temperature branch being the branch dominated by efficient free-free optically thick cooling, the intermediate branch being the branch where the radiation pressure dominates and the high temperature branch the slim branch where advection is the dominant source of cooling \citep{abramowicz1988}. Hence, there could also be a similar cycle, as with the ionization instability, where the disk oscillates between the optically thick branch and the optically thin advection dominated branch (\citealt{szuszkiewicz1998}, \citealt{lasota1991}). On the observational side, most LMXBs do not show such cycles though they reach luminosities where models are unstable (\citealt{nayakshin2000}, \citealt{done2007}). However, tentative evidence for a radiation-induced limit cycle comes from GRS 1915+105 when it goes super-Eddington (\citealt{belloni1997}, \citealt{done2004}). On the theoretical side, shearing-box simulations of radiation-dominated disks suggest that thin disks should indeed be thermally unstable (\citealt{jiang2013}). However, 3D radiative GRMHD simulations from \cite{sadowski2016} show that the radiation pressure instability could be suppressed by strong support of the azimuthal magnetic field as proposed in \cite{begelman2007}. It then remains unclear whether or not the inner parts of LMXBs should be thermally unstable. \\

We estimate here the radius at which the radiation pressure instability would occur, if it occurs. In LMXBs, opacity in the inner disk is dominated by Thomson scattering instead of free-free absorption. We estimate the unstable radius to be the radius where the radiation pressure is equal to the thermal pressure. Using $\kappa=\kappa_\mathrm{T}$ and the same method as in \S\ref{sec:analytical_model}, we find that the radiation pressure exceeds the gas pressure below the radius 
 \begin{equation}\label{eq:rad_thomson}
 r_\mathrm{rad}=20\alpha\left(\frac{q}{\beta}\right)^{-1}\dot{m}.
\end{equation}

If the radiation instability were to produce heating fronts propagating in the inner magnetized disk, as with the ionization instability, we would always observe QPOs once the disk reaches $\dot{m}\gtrsim0.01$ (for $\beta=10$), which is not the case. It might be that in a highly magnetized disk the radiation pressure instability is suppressed due to magnetic pressure support (\citealt{begelman2007}, \citealt{sadowski2016}). However, as the mass accretion rate increases the radiation pressure instability might leave the inner magnetized zone and perturb the outer disk. In our model, the transition radius between the magnetized disk and the unmagnetized disk goes as $\dot{m}^{-2/3}$ while the radius where the radiation pressure is triggered goes as $\dot{m}$. Then, during an eruption the two radii can cross when $\dot{m}$ rises. The value of $\dot{m}$ for which this happens depends on the amount of magnetic flux in the disk. Using Eq. (\ref{eq:rtr_marcel}) and (\ref{eq:rad_thomson}) we find 
\begin{equation}
\dot{m}_\mathrm{tr/rad}\approx10\alpha^{-3/5}\left(\frac{q}{\beta}\right)^{9/10}M_1^{-1}\beta^{3/10}.
\end{equation}
For $M_1=5$ and $\beta=10$, this gives $\dot{m}\approx 0.7$. This is close to, but slightly higher than, the value reached by LMXBs at the top of the eruption when Type B QPOs appear. This raises the interesting possibility that Type B QPOs might be related to the radiation pressure instability leaving the inner magnetized disk. Clearly this deserves further investigation and we leave a proper treatment of how the radiation pressure instability behaves in such a configuration to future work.

\section{Conclusions}\label{sec:conclusion}
We extended the work of \cite{scepi2020}, in which the authors reproduced the eruptions of DNe with a DIM model treating, for the first time, the transport of magnetic field. A strong feature of that model was that the magnetic field was advected inwards forming a highly magnetized region, that acts as a truncation of the inner disk, leaving the outer regions emptied of magnetic field. In the inner disk angular momentum is mostly transported by a magnetized outflow, while in the outer disk it is transported by turbulence as in an $\alpha$-disk. 

While \cite{scepi2020} focused on the long term dynamics of DNe, we focused here on the small scale behavior of the inner magnetized disk. Indeed, the modeled light curves show $\approx0.2$ mag quasi-periodic variations with a period of a few thousand seconds. We showed that these fluctuations originate from small eruptions in the inner disk. These eruptions are a manifestation of the thermal instability near the hydrogen ionization zone in the inner disk. 

We provided an analytical model of outflow-driven disks that we compared with classical $\alpha$-disks. Outflow-driven disks are colder and less dense than $\alpha$-disks and as a consequence the ionization instability is triggered at smaller radii than in an $\alpha$-disk. Moreover, the accretion time scale in an outflow-driven disk is much shorter than the viscous time scale. Altogether, this leads the ionization instability in the inner disk to have eruptions of much smaller amplitude and lower recurrence time scales than in a viscous disk. We showed that our analytical estimates reproduce well the results of the simulations of \cite{scepi2020} concerning the structure of the disk and the period of the fluctuations.

We also compared the period of the fluctuations to the period of observed QPOs of DNe. We find good agreement between our model and observations. Better constraints on the observed mass accretion rates as well as the magnetization would be needed to challenge our model. One strong prediction of our model is the fact that our QPOs should only be seen between $3\times10^{14}$ and $1.5\times10^{16}\:\mathrm{g\:s^{-1}}$, so in quiescence or at the beginning or end of an outburst. Out of the 7 observations that could be explained by our model 6 are indeed observed in quiescence or at the end of an outburst. Since it has been harder to observe DNe in quiescence, we might expect our QPOs to be more frequent than is presently thought. A continuous observation of the QPO frequency over an entire quiescent state until an outburst would test our model. Indeed, the QPO period should scale as $\dot{M}^{0.46}$ and disappear during the outburst. If such correlations were observed this would be a major step in our understanding of DNe as it would provide additional indirect proof of the existence of highly magnetized, outflow-driven disks in DNe.

Finally, we investigated the possibility that the ionization instability or the radiation pressure instability, in an inner magnetized disk, could explain some QPOs in LMXBs. The ionization instability could produce QPOs in LMXBs with frequencies of $\approx0.1$ mHz at the end of an outburst and higher frequencies deeper in quiescence. It cannot, however, explain any observed QPOs in LMXBs to date. We speculated that Type B QPOs might be a manifestation of the radiation pressure instability leaving the inner magnetized disk and reaching the outer disk. A verification of this idea is left to future work.

\section*{Acknowledgements}

NS thanks Guillaume Dubus and Geoffroy Lesur for fruitful discussions. We acknowledge support from NASA Astrophysics Theory Program grant NNX16AI40G.

\section*{Data Availability}
The data underlying this article will be shared on reasonable request to the corresponding author.




\bibliographystyle{mnras}
\bibliography{biblio} 


\bsp	
\label{lastpage}
\end{document}